\documentstyle[12pt,aaspp4,epsf]{article}

\slugcomment{Submitted to The Astronomical Journal}

\lefthead{D. Merritt}
\righthead{Optimal $N$-body Smoothing}

\begin{document}

\title{Optimal Smoothing for $N$-Body Codes}

\author{D. Merritt}
\affil{Department of Physics and Astronomy, Rutgers University,
    New Brunswick, NJ 08855}
\bigskip
\centerline{Rutgers Astrophysics Preprint Series No. 181}

\begin{abstract}
In any collisionless $N$-body code, there is an optimal choice for the 
smoothing parameter that minimizes the average error in the force 
evaluations.
We show how to compute the optimal softening length in a direct-summation
code and demonstrate that it varies roughly as $N^{-1/3}$.
\end{abstract}

\section{Introduction}

The role of smoothing in an $N$-body code depends on the 
nature of the system being modelled.
When simulating a system containing a modest number of point 
masses, like an open star cluster, the goal is to reproduce 
the exact level of graininess that exists in the real cluster.
Ideally no softening would be used, but a nonzero softening length 
is usually included to eliminate the divergent 
forces that would result from a close encounter between two ``stars.''
When modelling globular clusters, the softening length may be 
increased to reduce the graininess of the potential to the level 
expected in a system containing a larger number of particles than 
can easily be handled in the computer.
When simulating a system containing finite-sized objects, such as 
a galaxy cluster, the softening length may be increased still 
more to approximate the physical size of a typical object.

But $N$-body codes are often used to study the evolution of 
systems, like galaxies, that contain extremely large numbers of particles.
A code designed for such problems is sometimes called 
``collisionless,'' since the system being modelled 
has a two-body relaxation time that greatly exceeds the 
elapsed time of the simulation.
In a collisionless $N$-body code, the particles do not represent 
real objects; they are simply Monte-Carlo realizations of the 
underlying, smoother mass distribution.
The purpose of the smoothing in a collisionless code is to generate 
from the $N$ particles an accurate estimate, at each time step, 
of the gravitational forces corresponding to the much smoother 
system being modelled.

Smoothing parameters in $N$-body codes are often 
chosen in an {\it ad hoc} way.
Here we show that the softening length in a collisionless code
can be chosen objectively,
so as to minimize the average errors in the force determinations.
Too small a value for the smoothing parameter yields an estimate 
that is overly noisy, reflecting finite-$N$ fluctuations in 
the forces.
Too large a smoothing parameter reduces the noise but increases 
the error from the ``bias,'' i.e. the systematic 
misrepresentation of the force due to the failure to resolve real 
features with scale lengths less than the softening length.
In general, a unique value for the smoothing parameter may be found 
such that the combined error from these two sources is minimized.

The existence of an optimal softening length follows from the 
assumed smoothness of the underlying mass distribution; no such 
optimal degree of smoothing can be defined for a ``collisional'' 
code, where the accuracy of the force calculations can always be 
increased by decreasing the softening length.

Our analysis will focus on the simplest $N$-body algorithm, a 
direct-summation code in which the smoothing is imposed via a 
fixed softening length associated with each particle.
But the general principles carry over without significant change 
to codes that implement the smoothing in very different ways, 
e.g. via a grid, or via expansion of the potential in a truncated 
basis set.

\section{Definitions}

The gravitational force on particle $i$, $1\le i\le N$ 
in a direct-summation $N$-body code is given by
\begin{equation}
{\bf F}_i = Gm^2\sum_{j=1}^N{{\bf x}_j - {\bf 
x}_i\over\left(\epsilon^2 + |{\bf x}_i - {\bf 
x}_j|^2\right)^{3/2}},
\end{equation}
where $\epsilon$, the softening length, determines the degree 
of smoothing.
If the $N$ particles represent an underlying, smooth distribution 
of mass, there also exists a ``true'' value ${\bf 
F}_{true}({\bf x}_i)$ 
for the force on particle $i$.
For instance, if the $N$ positions are generated from a Plummer 
density profile, ${\bf F}_{true}$ is the gravitational force 
corresponding to the Plummer mass distribution.

The optimal choice of $\epsilon$ may be defined as the value that 
minimizes the average deviation between ${\bf F}_i$ and ${\bf 
F}_{true}({\bf x}_i)$.
There are many possible ways to define this average deviation; 
the simplest is the average square error, or
\begin{equation}
{\rm ASE} = {1\over N}\sum_{i=1}^N |{\bf F}_i - {\bf F}_{true}({\bf x}_i)|^2.
\end{equation}
We may treat ${\bf F}$ as a continuous function, defined via 
Eq. (1) at any point ${\bf x}$.
The continuous analog of the average square error is the 
integrated square error, or
\begin{equation}
{\rm ISE} = \int \rho({\bf x})|{\bf F}({\bf x}) - {\bf F}_{true}({\bf x})|^2 
d{\bf x}
\end{equation}
with $\rho({\bf x})$ the true density, normalized to unit total 
mass.

If we imagine generating many $N$-particle Monte-Carlo realizations 
of the same smooth model, we can define the expectation or mean 
value of the ISE as
\begin{equation}
{\rm MISE} = E({\rm ISE}) = 
E\int\rho({\bf x})|{\bf F}({\bf x}) - {\bf F}({\bf x})_{true}|^2 d{\bf x},
\end{equation}
where $E$ indicates an average over many realizations.
It is easy to show that the MISE contains contributions from two 
terms, representing the bias and the variance of the 
force estimates.
The bias is the mean deviation of the computed force at some 
point from the true force, $E{\bf F} - {\bf F}_{true}$; 
its integrated square value is
\begin{equation}
{\rm ISB} = \int\rho({\bf x})|E {\bf F}({\bf x}) - {\bf F}_{true}({\bf 
x})|^2 d{\bf x}.
\end{equation}
The variance is the mean square deviation of the force estimate 
from its mean value $E{\bf F}$; the integrated variance
is ${\rm IV} = {\rm MISE} - {\rm ISB}$.

The bias increases with the softening length while the variance 
falls off -- a greater amount of smoothing produces smaller average 
fluctuations in the forces but also tends to smooth over real
features of small scale.
For $N$-particle realizations of a given model, therefore, there is 
an 
optimal choice of softening length $\epsilon_{opt}$ such that the sum 
${\rm ISB} + {\rm IV}$ is minimized.
One typically finds in problems of this sort (e.g. Scott 1992, p. 131) 
that the contribution to the total error from the bias and from 
the variance are of the same order when the smoothing length is optimized.
Furthermore, the variance decreases with particle number while the bias is 
independent of $N$ (e.g. Silverman 1986, p. 39); thus as $N$ 
increases, one expects $\epsilon_{opt}$ to decrease.

\section{Two Examples}

We illustrate these ideas using $N$ equal-mass points generated 
from two simple models.
The first is a spherical Plummer model of unit scale length:
\begin{equation}
\rho(r)={3\over 4\pi} (1+r^2)^{-5/2}.
\end{equation}
The expectation value of the forces generated by a finite set of 
particles drawn from the density (6) will be spherically 
symmetric.
We may therefore compute the terms in Eqs. (4) and (5) on a linear 
grid in radius, taking averages over a large number of 
independent 3-D realizations of the particle distribution 
for any specified $N$.
The number of Monte-Carlo realizations was taken to be $N_{MC}=3\times 
10^6/N$, e.g. 1000 realizations for $N=3000$.

Fig. 1 shows how the contribution to the force errors from the 
bias and variance depend on $\epsilon$ for $N=1000$.
As expected, the bias increases with $\epsilon$ while the 
variance decreases, producing an optimal value of $\epsilon$ for which 
the MISE is minimized, $\epsilon_{opt}\approx 0.16$.
At $\epsilon=\epsilon_{opt}$, roughly 60\% of 
the total error comes from the variance and the remainder 
from the bias.

Fig. 2 shows how the MISE vs. $\epsilon$ curve changes with 
$N$.
As $N$ increases at fixed $\epsilon$, the variance decreases, 
roughly as $N^{-1}$.
The result is an optimal $\epsilon$ that decreases with $N$.
Fig. 3 presents $\epsilon_{opt}$ and MISE($\epsilon_{opt})$ 
as functions of $N$.
The two quantities vary roughly as power-laws in $N$, with
\begin{equation}
\epsilon_{opt}\approx 1.1\times N^{-0.28},\ \ \ \ {\rm MISE}_{opt}
\approx 0.21\times N^{-0.66}.
\end{equation}

Our second example is based on a spherical model with the density 
law
\begin{equation}
\rho(r) = {1\over 2\pi} r^{-1}(1+r)^{-3}
\end{equation}
which is a good approximation to the light distribution in 
early-type galaxies (Hernquist 1990).
Figure 3 shows $\epsilon_{opt}$ and MISE($\epsilon_{opt}$) as 
functions of $N$.
The relations in this case can be approximated as
\begin{equation}
\epsilon_{opt}\approx 1.5\times N^{-0.44},\ \ \ \ {\rm MISE}_{opt}
\approx 0.39\times N^{-0.58}.
\end{equation}
The variation of $\epsilon_{opt}$ with $N$ in both models 
is such that the number of particles within one softening 
volume is approximately independent of $N$.
However in both cases the dependence seems to be significantly
different than $N^{-1/3}$.

\section{Discussion}

Selection of the softening length in $N$-body codes is 
often made on the basis of the error due {\it either} to the bias {\it or} 
the variance of the force estimates.
For instance, $\epsilon$ may be chosen to be small enough that 
some feature like the core is resolved (bias), or large enough 
that two-body relaxation is not a problem (variance).
Our discussion shows that the total error in the estimated forces 
in a collisionless code is 
the sum of these two components, and that their contributions 
vary in opposite ways with $\epsilon$,
leading to a formally optimal 
choice of $\epsilon$ for any $N$.

The softening length in Eq. (1) may be interpreted as the
window width of a kernel that is convolved with the density before
computing the gravitational forces.
Smoothing in collisionless $N$-body codes is usually carried out 
in other ways than via a kernel, e.g. on a grid.
(Notable exceptions are the $N$-body studies of galaxies
based on the GRAPE hardware, which incorporates Eq. (1)
[Ebisuzaki et al. 1993].)
However, any smoothing algorithm may be interpreted in terms 
of a generalized kernel whose shape and width varies with 
position (Scott 1992, p. 155); in this sense, the analysis given 
here is general.
One could use similar arguments to compute the optimal 
cell size in a grid-based algorithm, or the optimal number of 
terms in a basis-function expansion.

It is well known that ``collisionless'' $N$-body codes often exhibit 
nearly as much relaxation as ``collisional'' codes (e.g. Hernquist \&
Barnes 1990).
The reason is that the variance in the force estimates -- which,
crudely speaking, 
is responsible for the relaxation -- cannot be 
arbitrarily reduced without introducing a bias, as shown by
Fig. 1.
The primary design goal for collisionless $N$-body experimenters 
(aside from speed) should 
be to smooth in such a way as to minimize the ISE 
of the force estimates at every time step, since by doing so they
will have effectively minimized the variance.

One way to decrease the ISE is to vary the softening length with position
(and hence with time).
The optimal way of doing this is well known when the quantity to 
be estimated is the density itself (Abramson 1982), but apparently 
nothing is known about the best way to vary $\epsilon$ with position 
when estimating the gravitational forces.
This would be a fruitful topic for further study.

One would like to be able to estimate the optimal smoothing parameter
directly from the $N$ positions themselves, without having to know 
$F_{true}$.
In a direct-summation code, automatic choice of $\epsilon$ would
presumably have to be based on a time-intensive bootstrap algorithm.
More rapid choice of the smoothing parameter could be made in a
code that represents the forces via sums of terms like 
$R_{lm}(r)Y_{lm}(\theta,\phi)$, with $Y_{lm}$ an angular harmonic.
The radial functions $R_{lm}$ may themselves be represented via 
basis-function expansions (e.g. Clutton-Brock 1973), but it seems 
to have escaped general notice that the job may be done more 
simply and efficiently using smoothing splines (Wahba 1990).
Standard, and efficient, routines for computing the optimal 
degree of smoothing for such splines are widely available (e.g. Green
\& Silverman 1994).
The result would be a code in which the radial dependence of the 
forces was computed with minimum ISE at every time step.

One sometimes sees implementations in which $N$ has been increased
to very large values without corresponding adjustments in the
degree of smoothing (e.g. Hernquist, Sigurdsson \& Bryan 1995).
Figure 2 shows that such a practice is likely to put the user far to
the right of the minimum in the MISE curve.

Finally, we note that objective criteria for comparing the 
performance of different $N$-body codes are generally lacking.
A natural criterion for collisionless codes would be the rate 
of decrease of the MISE with $N$.
An optimal potential solver 
could be defined as one for which the MISE decreased most rapidly 
with $N$. 

\bigskip
This work was supported by NSF grant AST 90-16515 and NASA grant 
NAG 5-2803.
I thank G. Quinlan for helpful discussions.

\clearpage

\begin{figure}
\plotone{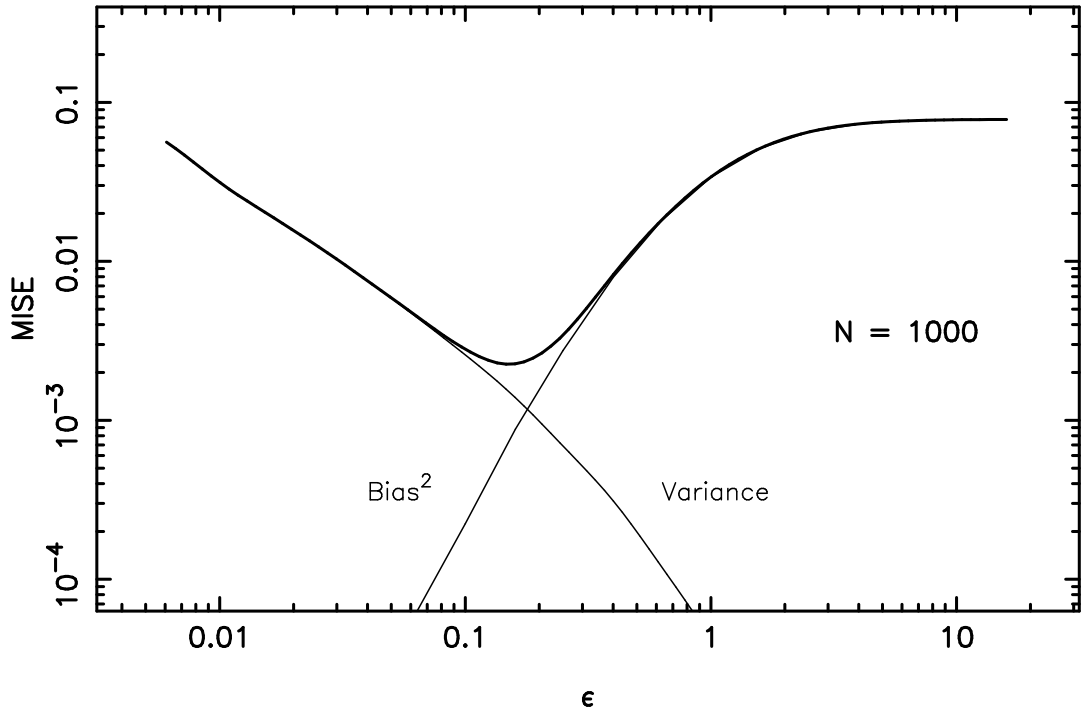}
\caption{Bias-variance tradeoff in 
the force evaluation errors for $N=1000$ realizations of a 
spherical Plummer model. \label{fig1}}
\end{figure}

\begin{figure}
\plotone{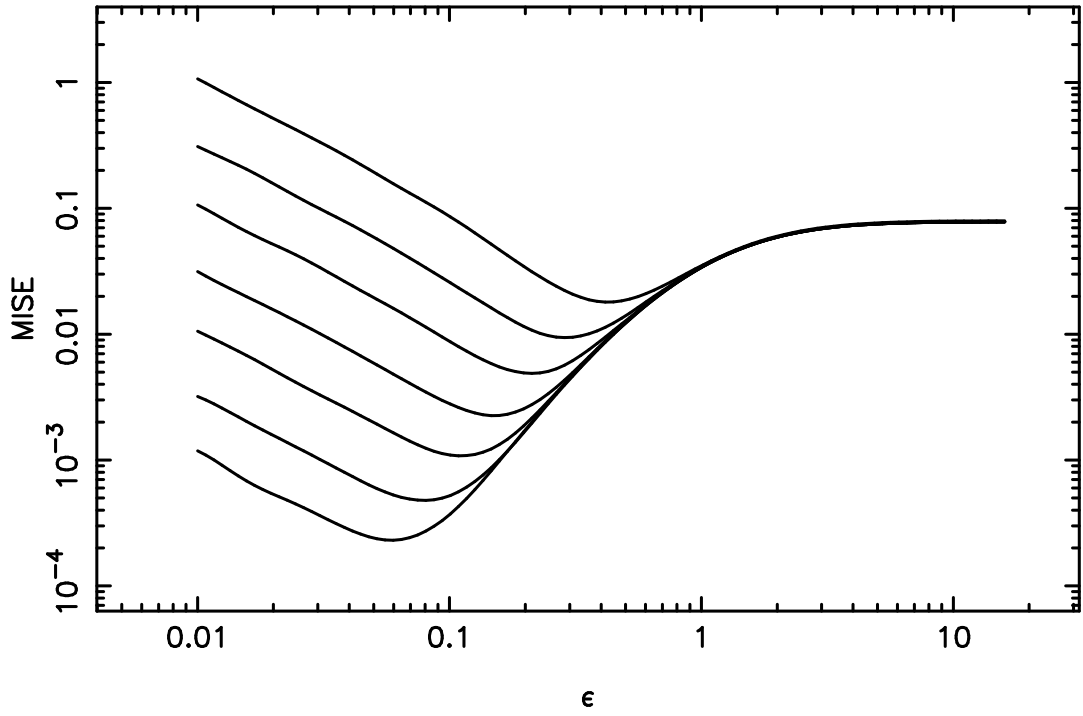}
\caption{MISE$(\epsilon)$ for $N=30, 
100, 300, 1000, 3000, 10000$ and $30000$; $N$ increases downward.
\label{fig2}}
\end{figure}

\begin{figure}
\plotone{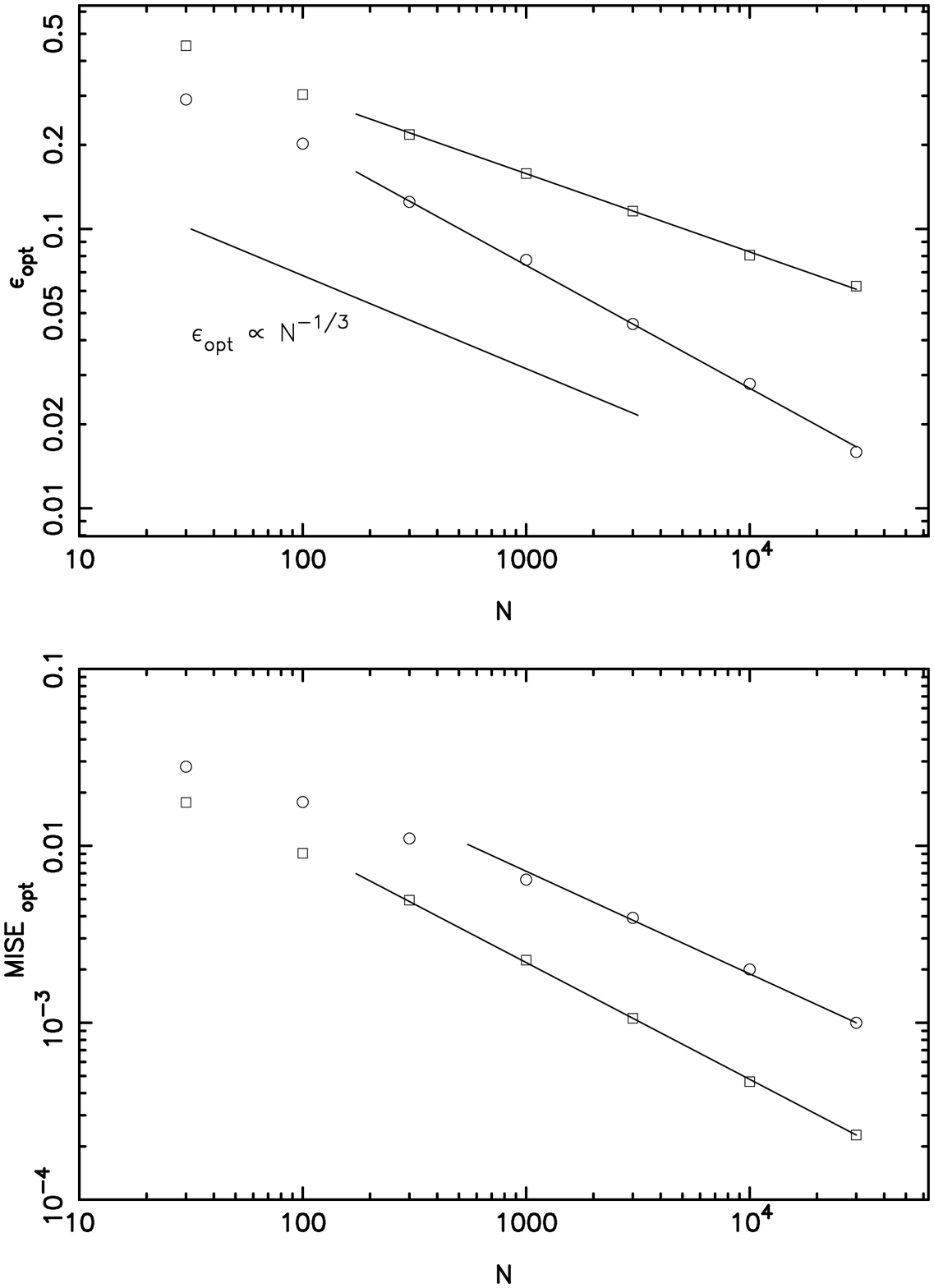}
\figcaption[figure3.ps]{\label{fig3}} Dependence of the optimal 
softening length $\epsilon_{opt}$, and the MISE at 
$\epsilon=\epsilon_{opt}$, on number of particles $N$ in two models:
Eq. (6) (squares) and Eq. (8) (circles).
The relations of Eqs. (7) and (9) are indicated by solid lines.
\end{figure}





%


\begin{thebibliography}{}
\bibitem[Abramson 1982]{abr82} Abramson, I. S. 1982, Ann. Statist., 10,
	1217
\bibitem[Clutton-Brock 1973]{clu73} Clutton-Brock, M. 1973, Astrophys.
	Sp. Sci., 23, 55
\bibitem[Ebisuzaki 1993]{ebi93} Ebisuzaki, T. et al. 1993, P. A. S. J., 
	45, 269
\bibitem[Green \& Silverman 1994]{gre94} Green, P. J. \& Silverman,
	B. W. 1994, Nonparametric Regression and Generalized
	Linear Models, London: Chapman \& Hall, ch. 8
\bibitem[Hernquist \& Barnes 1990]{her90} Hernquist, L. \& Barnes, J. E.
	1990, Ap. J., 349, 562
\bibitem[Hernquist 1990]{her90} Hernquist, L. 1990, Ap. J., 356, 359
\bibitem[Hernquist, Sigurdsson \& Bryan 1995]{her95} Hernquist, L.,
	Sigurdsson, S. \& Bryan, G. 1995, Ap. J., 446, 717
\bibitem[Scott 1992]{sco92} Scott, D. W. 1992, Multivariate Density 
	Estimation, New York: John Wiley \& Sons
\bibitem[Silverman 1986]{sil86} Silverman, B. W. 1986, Density 
	Estimation for Statistics and Data Analysis, London: Chapman \& 
	Hall
\bibitem[Wahba 1990]{wah90} Wahba, G. 1990, Spline Models for
	Observational Data, Philadelphia: SIAM

\end{thebibliography}
\end{document}